\documentstyle[12pt,aaspp4]{article}
\begin{document}

\title{A Search for Exozodiacal Dust and Faint Companions Near Sirius,
Procyon, and Altair with the NICMOS Coronagraph\altaffilmark{1}}
\author{Marc J. Kuchner}
\affil{Palomar Observatory, California
Institute of Technology, Pasadena, CA 91125}
\authoremail{mjk@gps.caltech.edu}
\author{Michael E. Brown\altaffilmark{2}}
\affil{Division of Geological and Planetary Sciences, California
Institute of Technology, Pasadena, CA 91125}
\altaffiltext{1}{Based on observations with the NASA/ESA Hubble Space Telescope,
obtained at the Space Telescope Science Institute, which is operated by the
Association of Universities for Research in Astronomy, Inc. under NASA
contract No. NAS5-26555.}
\altaffiltext{2}{Alfred P. Sloan Research Fellow}

\begin{abstract}

We observed Sirius, Altair, and Procyon with the NICMOS Coronagraph on
the Hubble Space Telescope to look for scattered light from exo-zodiacal
dust and faint companions within 10 AU from these stars.  We did not
achieve enough dynamic range to surpass the upper limits set by IRAS
on the amount of exo-zodiacal dust in these systems, but we did set
strong upper limits on the presence of nearby late-type and sub-stellar 
companions. 

\end{abstract}


\section{Introduction}

Several main sequence stars are close enough that a large
telescope operating at the diffraction limit can resolve the terrestrial
planet-forming region within 10 AU from the star \markcite{kuch98}
(Kuchner, Brown \& Koresko 1998).  We used the NICMOS coronagraph
to image three of the nearest main-sequence stars in the
near-infrared to look for circumstellar material---exozodiacal dust
and faint companions---in this relatively uncharted circumstellar
region.

Any dust orbiting close to one of our targets must have been generated
recently by some population of larger bodies, since dust near a
star quickly spirals into the star due to Poynting-Robertson drag
(Robertson 1937).  Ten micron diameter dust 3 AU from a G star spirals 
into the star on time scales of $\sim10^5$ years; for brighter stars,
this timescale is shorter.  Around the sun, zodiacal dust forms when
asteroids collide and when comets outgass; a search for exozodiacal
dust is therefore implicitly a search for extra-solar asteroid or comet-like bodies
that make dust. 

Several disks around nearby main-sequence stars appear to have
exozodiacal components.  Some, such as the disk associated with
Beta-Pictoris, reveal their warm dust as a silicate
emission feature at 10 microns \markcite{tele91}
(Telesco \& Knacke 1991).  Others, like the disk around HR 4796,
show resolved emission at 10 microns that is interpreted as exozodiacal
\markcite{koer98} (Koerner et al. 1998).  Dust clouds like these,
which have $\sim 1000$ times as much warm dust as our sun,
emit thermal radiation substantially in excess of the stellar
photospheric emission, and can often be detected photometrically
by studying the spectral energy distribution of the star in
the mid-infrared.  However, many less massive exozodiacal
clouds may never be detectable photometrically because no stellar
spectrum is known to better than $\sim$ 3\% in the mid-infrared \markcite{cohe96}(Cohen et al. 1996).  We have begun to search for disk
that are too faint to be detected photometrically by spatially
resolving the critical regions less than 10 AU from nearby stars.

Coronagraphic images can also reveal faint companions to nearby
stars.  Such companions can go undetected by radial velocity surveys
because of their small masses or long orbital periods. 
Siruis and Procyon both have white dwarf companions whose orbits
are well studied, but analyses of the orbital motion in these
systems leave room for additional low mass companions.

The Sirius system in particular, so prominent in the night sky, has
spurred much debate in the last century over its properties.  Three
analyses of the proper motion of Sirius have suggested that there may
be a perturbation in the orbit of Sirius B with a $\sim 6$ year
period \markcite{vole32}\markcite{walb83}\markcite{bene95}
(Volet 1932; Walbaum \& Duvent 1983; Benest \& Duvent 1995).
These analyses do not indicate whether the perturbing body orbits
Sirius A or B, and dynamical simulations indicate that stable
orbits exist around both Sirius A and B at circumstellar distances
up to more than half the binary's periastron separation
\markcite{bene89}(Benest 1989).  If such a companion
were in a simple face-on circular orbit it would appear at a
separation of 4.2 AU (1.6 arcsec) from Sirius A or a separation of
3.3 AU (1.3 arcsec) from Sirius B assuming that the masses for
Sirius A and B a are 2.1 and 1.04 M${}_{\odot}$ respectively
\markcite{gate78}(Gatewood \& Gatewood 1978).  Benest \& Duvent \markcite{bene95}(1995) do not derive a mass for the hypothetical
companion from observations of the system, but they estimate that
a perturber much more massive than 0.05 M${}_{\odot}$ would rapidly
destroy the binary.

Perhaps the most interesting debate about Sirius is whether or not
the system appeared red to ancient observers $\sim 2000$ years ago.
Babylonian, Graeco-Roman and Chinese texts from this time period have
separately been interpreted to say that Sirius was a red star
\markcite{brec79}\markcite{schl85}\markcite{bonn91}
(Brecher 1979; Schlosser \& Bergman 1985, Bonnet-Bidaud \& Gry 1991).
\markcite{tang86}\markcite{vang84}\markcite{mccl87}
Tang (1986), van Gent (1984) and McCluskey (1987) have attacked
some of these reports, claiming that they represent mistranslations or misidentifications of the star.  However, if the Sirius system did
indeed appear red, the existence of a third star in the group
interacting periodically with Sirius A could explain the effect
\markcite{bonn91}(Bonnet-Bidaud 1991).

The low mass companions ($< 0.1$ M${}_{\odot}$) that we could hope to
detect with NICMOS are late-type stars or warm brown dwarfs, shining with
their own thermal power in the near-infrared.  Schroeder and
Golimowski \markcite{schr96}(1996) recently imaged Sirius, Procyon and
Altair at visible wavelengths with the Planetary Camera on HST in a
search for faint companions.  Our observations  are more
sensitive to late-type companions because of the high dynamic range
of the NICMOS coronagraph, and because these objects are brighter in
the infrared than the optical.

\section{Observations}

We observed our target stars with the NICMOS Camera 2 coronagraph on
five dates during 1999 October.  We used the F110 filter, the
bluest available near-infrared filter, with an effective
wavelength of 1.104 microns, to take advantage of the higher
dynamic range the coronagraph has at shorter wavelengths.  We took
images of Sirius and Procyon at two different orientations,
rolling the telescope about the axis to the star
by $15^{\circ}$ between them.  When we searched for faint
companions in the images, we subtracted the images taken at one
roll angle from the images taken at the other angle to cancel
the light in the wings from the image of the occulted star.  We
planned to image Altair at a second roll angle, but on
our second visit to the star the telescope's Fine Guidance Sensors
failed to achieve fine lock on the guide star due to ``walkdown''
failure.

At each roll angle we took 50 short exposures in ACCUM mode,
lasting 0.6 seconds each, and we co-added them, for total integration
times of 30 seconds.  Even though we used the shortest available
exposure times, our images saturated interior to about 1.9
arcseconds for Sirius, 1.4 arcseconds for Procyon, and 0.7
arcseconds for Altair.  The actual coronagraphic hole is only 0.3
arcseconds in radius.  Table 1 summarizes the timing and orientations
of our observations.

Figure 1 shows an image of Sirius taken at one roll angle.  The
white dwarf Sirius B appears to the left of Sirius A,
at a separation of 3.79 arcseconds.  We derived photometry of
Sirius B at 1.1 microns from the roll subtracted image of Sirius
using a prescription from \markcite{riek99} Rieke (1999).  We
measured the flux in circular apertures with radii 7.5 pixels
around the positive and negative images and multiplied the flux
in those regions by an aperture correction of 1.110 to extrapolate
to the total flux.  Then we used a factor of $1.996 \times 10^{-6}$
Jy/ADU/Sec to convert from ADU to Janskys.  In this manner, we
measured the flux in Sirius B to be $0.503 \pm 0.15$ Jy.  Procyon
also has a white dwarf companion, Procyon B, that has been
previously detected by HST  \markcite{prov97}(Provencal et al. 1997).
It is not visible in our images, because it is currently at a
separation of $\sim5$ arcseconds from Procyon A.

\section{Exozodiacal Dust}

We compared our observations of Sirius, Procyon and Altair
to a simple model for what our zodiacal cloud would look like if it
were placed aound these stars.  Kelsall et al. \markcite{kels98}(1998)
fit an 88-parameter model of the zodiacal cloud to the maps of the
infrared sky made by the Diffuse Infrared Background Experiment (DIRBE)
aboard the Cosmic Background Explorer (COBE) satellite.
We used the smooth component of this model, which has a face-on
optical depth of $7.11 \times 10^{-8} (r / 1 {\rm AU})^{-0.34}$,
and extrapolated it to an outer radius of 10 AU.  We used a
scattering phase function consisting of a linear combination of three
Henyey-Greenstein functions that Hong \markcite{hong85}(1985) fit
to visible light observations of the zodiacal cloud with the
Helios Satellite, and we assumed an albedo of 0.2, from the Kelsall et
al. \markcite{kels98}(1998) fit to the 1.25 micron DIRBE maps.

For this part of our search, we could not use roll-subtraction to
cancel the light in the images of our target stars, because this
approach would also cancel most of the light from an exozodiacal
disk, even if the disk were edge-on.  Instead we subtracted images
of Altair from the images of Procyon and Sirius,
with the assumption that all three of our stars would not have identical circumstellar structures.  We used the IDP3 data analysis software 
\markcite{idp3}(Lytle et al. 1999) to perform sub-pixel shifts on
the images of Altair before we subtracted them from our images of
Sirius and Procyon to compensate for the slightly different
relative alignments of the three stars and the coronagraphic hole. 

Figures 2a and 3a show our images of Sirius and Procyon minus our
image of Altair.  Software masks hide the regions where the images
are saturated and the four main diffraction spikes. The bright horn
just above the masked area in the Procyon image is a well known
NICMOS artifact.  

Figures 2b and 3b show the same images plus synthesized images
of exozodiacal clouds seen in scattered light.  The
models are brightest immmediately to the left and the right of the
circular masked regions.  The symmetry planes of the model disks
are inclined $30^{\circ}$ from edge-on.  The dust densities in these
have been enhanced to $ > 10^5 \times$ solar levels so
they are marginally discernible from the residuals from the PSF
subtraction.  We used these models for the sake of comparison with the
solar zodiacal cloud; real disks with this much dust would be
severely collisionally depleted in their centers, unlike the solar cloud.
Despite the high dynamic range of the NICMOS coronagraph and our
efforts at PSF calibration, we were not able to improve upon photometric
detection limits for exozodiacal dust around these stars;  if the stars
actually had this much circumstellar dust, the thermal emission from
the dust would have been seen as a photometric excess by IRAS.

Our study demonstrates the difficulty of detecting exozodiacal dust
in the presence of scattered light from a bright star in a single-dish
telescope.  Faint companions can be differentiated from the wings of
the telescope PSF by techniques like roll subtraction, but if
exozodiacal clouds resemble the solar zodiacal cloud, light from
these clouds will resemble the PSF wings.  Even though coronagraphs
can suppress the PSF wings from an on-axis source by as much as an
order of magnitude, the dynamic range obtainable with a coronagraph
on a large, diffraction-limited telescope in the near-infrared is far
from that required to probe dust levels comparable to the solar cloud.

\section{Faint Companions}

For our faint companion search we created roll-subtracted images
of Sirius and Procyon using the IDP3 software.
To find the detection limits for faint companions among the non-gaussian
PSF residuals, we tested our abilities to see artificial stars
added to our images.  We examined roughly 350 copies of
the PSF-subtracted images of each of Sirius, Procyon and Altair
with help from a few of our patient colleagues.
To five-sixths of the images, we added images of artificial stars,
copied from our image of Sirius B, at random positions and magnitudes that were unknown to the examiner.  The other images were left unaltered and mixed with the images that contained artificial stars.  The examiners were shown each
image one at at time, and asked whether they could say confidently
that the image they were shown had an artificial star.
Only 2\% of the time did an examiner claim to see an artificial star
when none had been added to the image.
We quote as our detection limit the threshold for finding 90\% of
the artificial companions; that is, the examiners reported 90\% of
the artificial companions brighter than our detection limit at a given separation.  Figure 4 shows these detection limits.  For
comparison, we plot the expected magnitudes of two kinds of
possible companions to these objects:
an L0 dwarf like 2MASP J0345432+254023 \markcite{kirk99}
(Kirkpatrick et al. 1999) and a cool brown dwarf, Gl229B
\markcite{matt96}(Matthews et al. 1996).

Our apparent detection limits for Procyon are somewhat better than our
detection limits for Sirius because Procyon is almost a magnitude fainter
in the the near infrared; the two sets of observations yielded about the
same dynamic range.  Although Altair is fainter than Procyon, our
absolute detection limits for faint companions to Altair are not much
better than our detection limits for companions around Procyon because
we have only exposures at only one roll angle for Altair.  Based on
figure 4, we can rule out the existence of M dwarf companions farther
than $1.4$ arcseconds from Altair, $1.6$ arcseconds from Procyon and $1.8$ arcsecons from Sirius at greater than the 90\% confidence level.  It should be noted, however, that the coronagraph hole is only 3.5 arcseconds
from the edge of the chip, and that artificial faint companions that
were behind one of the four main diffration spikes at one roll angle
were harder to detect than artificial companions at other position
angles; figure 4 is averaged over position angle.  If we compare our upper limits to the J magnitude of T Dwarf Gl229B \markcite{matt96}(Matthews et al. 1996), we find that we can rule out dwarfs hotter than this object---including all L dwarfs---farther than  $2.3$ arcseconds from Procyon and $\sim 3.0$ arcseconds from Sirius and Altair.  For comparison, note that Gl 229B was discovered 7.7 arcseconds from a M1V star with an intrinsic luminosity 5 magnitudes fainter than Sirius in
the J band \markcite{naka95} (Nakajima et al. 1995).


We do not see any evidence for previously undetected faint companions
in our images.  If there were a low-mass companion orbiting Sirius at
4.2 AU we could not detect it because it would lie in the saturated parts of
our images.  However, we did survey a large fraction of the space where a
companion orbiting Sirius B might be found, and we could have detected a
brown dwarf like Gl 229B throughout most of this zone.  If there is a third
object in the Sirius system, and it orbits Sirius B with a 6 year orbit,
it is probably fainter than a brown dwarf.

\acknowledgments

We thank Glen Schneider and Aaron Evans for help with the data reduction,
and Edo Berger, John Carpenter, Micol Christopher, Ulyana Dyudina, David Frayer, Roy Gal, Pensri Ho, Matthew Hunt, Shardha Jogee, Olga Kuchner, Charlie Qi, Michael Santos, Alice Shapley and David Vakil for searching for artificial stars in our data.

This research made use of the Simbad database, operated at the Centre
de Donnees de Strasbourg (CDS), Strasbourg, France.
 
Support for this work was provided by NASA through grant number GO-07441.01-96A
from the Space Telescope Science Institute, which is operated by AURA,
Inc., under NASA contract NAS5-26555.

\newpage

\begin{deluxetable}{cccccc}
\footnotesize
\tablewidth{0 in}
\tablenum{1}
\tablecaption{\label{tbl-1}}
\tablecaption{Observations}
\tablehead{\colhead{Target} & \colhead{Spectral Type} & \colhead{Distance (pc)\tablenotemark{a}} & J \tablenotemark{b} & \colhead{UT Date} & \colhead{Orientation} 
} 
\startdata

Sirius  & A1V  & 2.64 & -1.34 & October 20 & 64.51${}^{\circ}$  \nl
        &      &      &       & October 22 & 81.51${}^{\circ}$  \nl
Procyon & F5IV & 3.50 & -0.40 & October 9  & 40.51${}^{\circ}$  \nl
        &      &      &       & October 21 & 55.51${}^{\circ}$  \nl
Altair  & A7V  & 5.14 &  0.39 &October 14 & -126.24${}^{\circ}$  \nl

\enddata

\tablenotetext{a}{From the Hipparcos Catalogue (Perryman et al. 1997)}
\tablenotetext{b}{From the SIMBAD online database}

\end{deluxetable}

\newpage

\figcaption{A coronagraphic image of the Sirius system.  The white dwarf
Sirius B, appears to the left of the residual light from Sirius A.  
Even though we used the shortest available exposure time, the
region $< 1.9$ arcsec from Sirius A is saturated.  \label{fig1}}  

\figcaption{a) An image of Sirius made using our coronagraphic image
of Altair to cancel the wings of the occulted PSF.  The saturated
regions of the image are hiden with a software mask.  b) The same image
plus a model of the scattered light from an exozodiacal cloud similar
to the solar zodiacal cloud but $2.5 \times 10^5$ times as bright.    \label{fig2}}
 
\figcaption{a) An image of Procyon using Altair as a PSF calibrator.
b) The same image plus a model of the scattered light for an exo-zodiacal cloud $7 \times 10^5$ times as bright as our own zodiacal cloud. \label{fig3}} 

\figcaption{Detection limits for faint companions around our three
target stars as a function of separation from the stars.  The magnitudes
of some representative cool objects, GL 229B and an L0 dwarf, are
shown for comparison.   \label{fig4}}

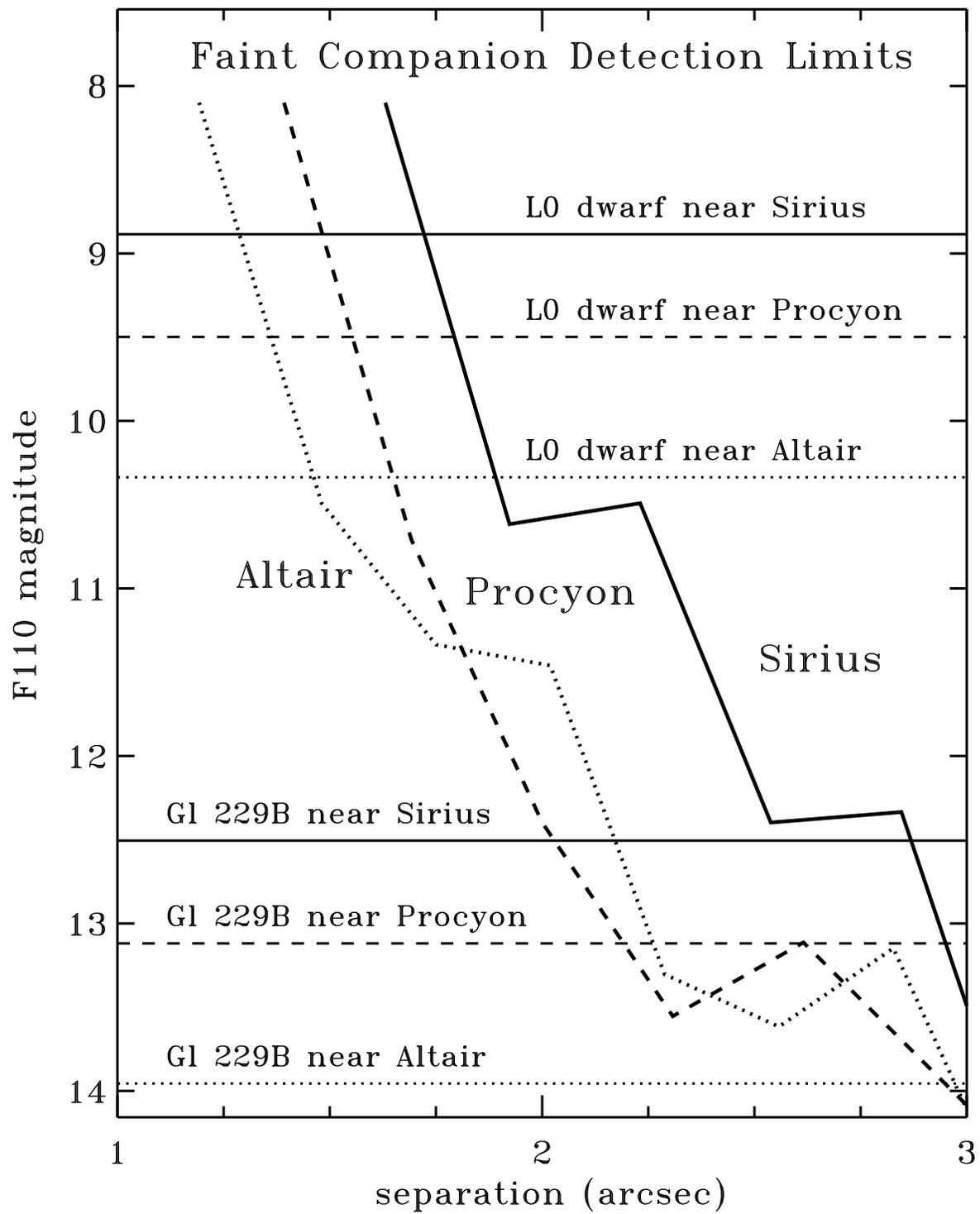
\end{document}